\documentclass[aps,prc,twocolumn,showpacs,showkeys,nofootinbib,superscriptaddress]{revtex4-2}

\usepackage{lipsum}

\usepackage{amsmath}  
\usepackage{amsfonts}
\usepackage{amssymb}
\usepackage{natbib} 
\usepackage{setspace}
\usepackage{graphicx} 
\usepackage{xspace}
\usepackage{cancel}
\usepackage{url}
\usepackage{color}

\usepackage{longtable}
\usepackage{graphicx}
\usepackage{amsmath}
\usepackage{amsfonts}
\usepackage{amssymb}
\usepackage{natbib} 
\usepackage{setspace}
\usepackage{xspace}
\usepackage{cancel}
\usepackage{physics}
\usepackage{tensor,braket,color,bbm,slashed}

\usepackage{newtxtext,newtxmath}
\usepackage{mathrsfs}

\def\K0bar{\overline{K^0}}
\def\bge{\begin{equation}}
\def\ene{\end{equation}}
\def\bg{\begin{eqnarray}}
\def\en{\end{eqnarray}}
\def\nn{\nonumber}

\def\qbar{{\overline{q}}}

\def\d0bar{{\bar{D}^0}}

\def\bm{\boldmath}

\def\bm{\boldmath}

\def\qbar{\overline{q}}
\def\Qbar{\overline{Q}}

\newcommand{\be}{\begin{equation}}
\newcommand{\ee}{\end{equation}}

\begin{document}

\title{
\begin{flushright}
\rightline{\hfill LFTC-19-13/51} 
\end{flushright}
Hadron properties in a nuclear medium and effective nuclear force from quarks:\\  
the quark-meson coupling model
}

\author{K.~Tsushima}
\affiliation{Laborat\'orio de F\'isica Te\'orica e Computacional, Universidade Cruzeiro 
  do Sul / Universidade Cidade de Sao Paulo, 01506-000, S\~ao Paolo, SP, Brazil}





\begin{abstract}
We give a short review of the quark-meson coupling (QMC) model, the quark-based model 
of finite nuclei and hadron interactions in a nuclear medium, 
highlighting on the relationship with the Skyrme effective nuclear forces. 
The model is based on a mean field description of nonoverlapping nucleon 
MIT bags bound by the self-consistent exchange of Lorentz-scalar-isoscalar, 
Lorentz-vector-isoscalar, and Lorentz-vector-isovector meson fields  
directly coupled to the light quarks up and down.
In conventional nuclear physics the Skyrme effective forces are very popular, but, 
there is no satisfactory interpretation of the parameters appearing in the Skyrme forces. 
Comparing a many-body Hamiltonian generated by the QMC model in the zero-range 
limit with that of the Skyrme force, it is possible to obtain 
a remarkable agreement between the Skyrme force and the QMC effective interaction.  
Furthermore, it is shown that 3-body and higher order N-body forces are naturally 
included in the QMC-generated effective interaction. 
\end{abstract}

\maketitle

\section{Introduction}
\label{intro}

This article intends to give a short review of the quark-meson coupling (QMC) 
model~\cite{Guichon:1987jp}, the quark-based model of 
finite nuclei and hadron properties in a nuclear medium.
Aside from the model basics, we highlight on the relationship 
with the Skyrme effective nuclear forces. 
(For detailed reviews of the QMC model, 
see Refs.~\cite{Saito:2005rv,Krein:2017usp,Guichon:2018uew}.)
The QMC model has been successfully applied to various studies   
of the properties of finite (hyper)nuclei~\cite{Guichon:1995ue,Saito:1996sf,
Saito:1996yb,Stone:2016qmi,Guichon:2018uew,Tsushima:1997cu,
Tsushima:1997rd,Guichon:2008zz,Tsushima:2002sm,Tsushima:2002ua,
Tsushima:2003dd}, hadron properties 
in a nuclear medium~\cite{Saito:1997ae,Tsushima:1997df,
Tsushima:1998qw,Tsushima:1998ru,Sibirtsev:1999jr,Tsushima:2002cc}, 
reactions involving nuclear targets~\cite{Sibirtsev:1999js,Shyam:2008ny,
Tsushima:2009zh,Shyam:2011aa,Chatterjee:2012ja,Tsushima:2012pt,Shyam:2016bzq,
Shyam:2016uxa,Shyam:2018iws}, and neutron star 
structure~\cite{RikovskaStone:2006ta,Whittenbury:2013wma,Thomas:2013sea}.
Self-consistent exchange of Lorentz-scalar-isoscalar ($\sigma$),  
Lorentz-vector-isoscalar ($\omega$), and Lorentz-vector-isovector ($\rho$) mean fields   
directly coupled only to the light quarks up and down, 
is the key feature of the model for achieving the novel saturation 
properties of nuclear matter, despite of its simplicity.  
All the relevant coupling constants for the $\sigma$-light-quarks,  
$\omega$-light-quarks, and $\rho$-light-quarks in any hadrons, 
are the same as those in nucleon, and they are fixed/constrained by the nuclear 
matter saturation properties. 
The physics behind this picture is the fact 
that the light-quark chiral condensates change faster than those of 
the strange and heavier quarks as nuclear density increases.  
The light-quark chiral condensates are the order parameters  
for chiral symmetry in QCD, and change in their 
magnitudes are one of the most important driving forces for 
partial restoration of chiral symmetry in a nuclear medium.
This is modeled in the QMC model by the approximation that  
the $\sigma$, $\omega$, and $\rho$ fields couple directly
only to the light quarks.

\section{Finite nucleus in the QMC model}
\label{qmc}

The description below is based on Refs.~\cite{Saito:2005rv,Krein:2017usp,Tsushima:2018goq}.
Although a Hartree-Fock treatment is possible within the QMC model~\cite{Krein:1998vc}, 
the main features of the results, especially the density dependence of 
nuclear matter energy density, is nearly identical to that of 
the Hartree approximation. Then, it is sufficient to discuss the Hartree 
approximation.
(See e.g., Ref.~\cite{Whittenbury:2013wma} for a neutron star structure studied by  
the Hartree-Fock approximation in the QMC model.)

Before explaining nuclear matter in the QMC model, 
we start with a finite nucleus. Using the Born-Oppenheimer approximation, 
a relativistic Lagrangian density, which gives the same mean-field equations
of motion for a finite (hyper)nucleus, is  
given~\cite{Saito:2005rv,Krein:2017usp,Tsushima:1997cu} below, 
where the quasi-particles moving in single-particle orbits are three-quark 
clusters with the quantum numbers of a nucleon, strange, charm or bottom hyperon 
when expanded to the same order in 
velocity~\cite{Guichon:1995ue,Saito:1996sf,Tsushima:1997cu,Tsushima:2002ua,
Tsushima:2003dd,Tsushima:2002cc}:  
\begin{eqnarray}
{\cal L}_{QMC} &=& {\cal L}^N_{QMC} + {\cal L}^Y_{QMC},
\label{eq:LagYQMC} \\
{\cal L}^N_{QMC} &\equiv&  \overline{\psi}_N(\vec{r})
[ 
i \gamma \cdot \partial - m_N^*(\sigma) \nn \\ 
& & \hspace{-8ex} - (\, g_\omega \omega(\vec{r})
+ g_\rho \dfrac{\tau^N_3}{2} b(\vec{r})
+ \dfrac{e}{2} (1+\tau^N_3) A(\vec{r}) \,) \gamma_0
] \psi_N(\vec{r}) \quad \nn \\
& & \hspace{-8ex} - \dfrac{1}{2}[ (\nabla \sigma(\vec{r}))^2 +
m_{\sigma}^2 \sigma(\vec{r})^2 ]
+ \dfrac{1}{2}[ (\nabla \omega(\vec{r}))^2 + m_{\omega}^2
\omega(\vec{r})^2 ] \nn \\
& & \hspace{-8ex} + \dfrac{1}{2}[ (\nabla b(\vec{r}))^2 + m_{\rho}^2 b(\vec{r})^2 ]
+ \dfrac{1}{2} (\nabla A(\vec{r}))^2, \label{eq:LagN} \\
{\cal L}^Y_{QMC} &\equiv&
\overline{\psi}_Y(\vec{r})
[ i \gamma \cdot \partial
- m_Y^*(\sigma) \nn \\
& & \hspace{-8ex} - (\, g^Y_\omega \omega(\vec{r})
+ g^Y_\rho I^Y_3 b(\vec{r})
+ e Q_Y A(\vec{r}) \,) \gamma_0
] \psi_Y(\vec{r}), 
\nn\\
& &\hspace{-8ex} (Y = \Lambda,\Sigma^{0,\pm},\Xi^{0,-},
\Lambda^+_c,\Sigma_c^{0,+,++},\Xi_c^{0,+},\Lambda_b,\Sigma_b^{0,\pm},\Xi_b^{0,-}).
\label{eq:LagY}
\end{eqnarray}
For a normal nucleus, ${\cal L}^Y_{QMC}$ in Eq.~(\ref{eq:LagYQMC}), 
namely Eq.~(\ref{eq:LagY}) is not needed.
In the above $\psi_N(\vec{r})$ and $\psi_Y(\vec{r})$
are respectively the nucleon and hyperon (strange, charm or bottom baryon) fields. 
The mean-meson fields represented by, $\sigma, \omega$ and $b$, which 
directly couple to the light quarks self-consistently, are  
the Lorentz-scalar-isoscalar, Lorentz-vector-isoscalar and the third component of  
Lorentz-vector-isovector fields, respectively, while $A$ stands for the Coulomb field.
They are defined by the mean expectations by, 
$\sigma(\vec{r}) = <\sigma(\vec{r})>$, $\omega(\vec{r}) = \delta^{\mu,0} <\omega^\mu(\vec{r})>$, 
and 
$b(\vec{r}) = \delta^{\mu,0}\delta^{i,3} <\rho^{\mu,i}(\vec{r})>$.

In the approximation that the $\sigma$, $\omega$ and $\rho$ fields couple
only to the $u$ and $d$ light quarks, the coupling constants for the hyperon 
appearing in Eq.~(\ref{eq:LagY}) are obtained/identified 
as $g^Y_\omega = (n_q/3) g_\omega$, and $g^Y_\rho \equiv g_\rho = g_\rho^q$, 
with $n_q$ being the total number of valence light quarks in the hyperon $Y$, 
where $g_\omega = 3 g^q_\omega$ and $g_\rho$ are 
the $\omega$-$N$ and $\rho$-$N$ coupling constants. $I^Y_3$ and $Q_Y$
are the third component of the hyperon isospin operator and its electric
charge in units of the positron charge, $e$, respectively.

The field dependent $\sigma$-$N$ and $\sigma$-$Y$
coupling strengths respectively for the nucleon $N$ and hyperon $Y$,  
$g^N_\sigma(\sigma)$ and  $g^Y_\sigma(\sigma)$, are implicitly  
in Eqs.~(\ref{eq:LagN}) and~(\ref{eq:LagY}), and defined by
\bg
& &m_N^*(\sigma) \equiv m_N - g^N_\sigma(\sigma)
\sigma(\vec{r}),  
\label{effnmass}
\\
& &m_Y^*(\sigma) \equiv m_Y - g^Y_\sigma(\sigma)
\sigma(\vec{r}),
\label{effymass}
\\
& & \hspace{3ex} (Y = \Lambda,\Sigma,\Xi, 
\Lambda_c,\Sigma_c,\Xi_c,\Lambda_b,\Sigma_b,\Xi_b), \nn
\en
where $m_N$ ($m_Y$) is the free nucleon (hyperon) mass. 
The dependence of these coupling strengths on the applied
scalar field ($\sigma$) must be calculated self-consistently within the quark
model~\cite{Guichon:1987jp,Guichon:1995ue,Tsushima:1997cu,Tsushima:2002ua,
Tsushima:2002sm,Tsushima:2002cc}.
Hence, unlike quantum hadrodynamics (QHD)~\cite{Walecka:1974qa,Serot:1984ey}, even though
$g^Y_\sigma(\sigma) / g_\sigma^N(\sigma)$ may be
2/3 or 1/3 depending on the number of light quarks $n_q$ in the hyperon 
in free space, $\sigma = 0$ (even this is true only when their bag 
radii in free space are exactly equal in the QMC model using the MIT bag), 
this will not necessarily be the case in a nuclear medium. 
We define $g^{N,Y}_\sigma \equiv g^{N,Y}_\sigma (\sigma = 0)$ for later convenience.
Note that, we will write explicitly the $\sigma$ dependence  
as $g^{N,Y}_\sigma (\sigma)$. Therefore, without the $\sigma$ dependence, 
$g^{N,Y}_\sigma$ are the coupling constants when $\sigma = 0$ in this article.
(The explicit expression will be given by Eq.~(\ref{sigmacc}).)

The Lagrangian density Eq.~(\ref{eq:LagYQMC}) [or Eqs.~(\ref{eq:LagN}) and (\ref{eq:LagY})] 
leads [lead] to a set of equations of motion for the finite (hyper)nuclear system:
\begin{eqnarray}
& &[i\gamma \cdot \partial -m^*_N(\sigma) - (\, g_\omega \omega(\vec{r}) \nn\\
& & \hspace{1ex} + g_\rho \dfrac{\tau^N_3}{2} b(\vec{r})
 + \dfrac{e}{2} (1+\tau^N_3) A(\vec{r}) \,)
\gamma_0 ] \psi_N(\vec{r}) = 0, \label{eqdiracn}\\
& &[i\gamma \cdot \partial - m^*_Y(\sigma)-
(\, g^Y_\omega \omega(\vec{r}) \nn\\ 
& & \hspace{1ex} + g_\rho I^Y_3 b(\vec{r})
+ e Q_Y A(\vec{r}) \,)
\gamma_0 ] \psi_Y(\vec{r}) = 0, \label{eqdiracy}\\
& &(-\nabla^2_r+m^2_\sigma)\sigma(\vec{r}) \nn \\
& & \hspace{3ex} =- \left[\dfrac{d m_N^*(\sigma)}{d \sigma}\right]\rho_s(\vec{r})
- \left[\dfrac{d m_Y^*(\sigma)}{d \sigma}\right]\rho^Y_s(\vec{r}),
\nn \\
& & \hspace{3ex} \equiv g^N_\sigma C_N(\sigma) \rho_s(\vec{r})
    + g^Y_\sigma C_Y(\sigma) \rho^Y_s(\vec{r}) , \label{eqsigma}\\
& &(-\nabla^2_r+m^2_\omega) \omega(\vec{r}) =
g_\omega \rho_B(\vec{r}) + g^Y_\omega
\rho^Y_B(\vec{r}) ,\label{eqomega}\\
& &(-\nabla^2_r+m^2_\rho) b(\vec{r}) =
\dfrac{g_\rho}{2}\rho_3(\vec{r}) + g^Y_\rho I^Y_3 \rho^Y_B(\vec{r}),
 \label{eqrho}\\
& &(-\nabla^2_r) A(\vec{r}) =
e \rho_p(\vec{r})
+ e Q_Y \rho^Y_B(\vec{r}) ,\label{eqcoulomb}
\end{eqnarray}
where, $\rho_s(\vec{r})$ ($\rho^Y_s(\vec{r})$), $\rho_B(\vec{r})=\rho_p(\vec{r})+\rho_n(\vec{r})$
($\rho^Y_B(\vec{r})$), $\rho_3(\vec{r})=\rho_p(\vec{r})-\rho_n(\vec{r})$, 
$\rho_p(\vec{r})$ and $\rho_n(\vec{r})$ are the nucleon (hyperon) scalar, 
nucleon (hyperon) baryon, third component of isovector,
proton and neutron densities at the position $\vec{r}$ in
the (hyper)nucleus. Notice that the terms on the right hand side of Eq.~(\ref{eqsigma}),
$- [{d m_N^*(\sigma)}/{d \sigma}] \equiv
g^N_\sigma\, C_N(\sigma)$ and
$- [{d m_Y^*(\sigma)}/{d \sigma}] \equiv
g^Y_\sigma\, C_Y(\sigma)$. 
(Recall $g^N_\sigma = g^N_\sigma (\sigma=0)$ and
$g^Y_\sigma = g^Y_\sigma (\sigma=0)$.)
At the hadronic level, the entire information
of the quark dynamics is condensed in the effective couplings
$C_{N,Y}(\sigma)$ of Eq.~(\ref{eqsigma}), which characterize the  
features of the QMC model, namely, {\it the scalar polarisability}. 
Furthermore, when $C_{N,Y}(\sigma) = 1$, which correspond to
a structureless nucleon or hyperon, the equations of motion
given by Eqs.~(\ref{eqdiracn})-(\ref{eqcoulomb}) 
can be identified with those derived from naive QHD~\cite{Walecka:1974qa,Serot:1984ey}.

The effective mass of hadron $h$ (in the present case nucleon and   
hyperon), will be calculated by Eq.~(\ref{hmass}).
The explicit expressions for 
$C_{N,Y}(\sigma) \equiv S_{N,Y}(\sigma) / S_{N,Y}(\sigma=0)$ 
is defined next, and the effective masses $m^*_{N,Y}$ are related by,
\bg
\dfrac{d m_{N,Y}^*(\sigma)}{d \sigma}
&=& - n_q g_{\sigma}^q \int_{bag} d^3y\, 
{\overline \psi}_q(\vec{y}) \psi_q(\vec{y})
\nn\\
& & \hspace{-5ex} \equiv- n_q g_\sigma^q S_{N,Y}(\sigma) \nn \\
& & \hspace{-5ex} = - \left[ n_q g_\sigma^q S_{N,Y} (\sigma=0)\right]
\left( \dfrac{S_{N,Y}(\sigma)}{\left[ n_q g_\sigma^q S_{N,Y} (\sigma=0)\right]} \right) \nn \\
& & \hspace{-5ex} \equiv - \left[ n_q g_\sigma^q S_{N,Y} (\sigma=0)\right]\, C_{N,Y}(\sigma) \nn \\
& & \hspace{-5ex} \equiv - \dfrac{d}{\partial \sigma}
\left[ g^{N,Y}_\sigma(\sigma) \sigma \right],
\label{Ssigma}
\en
where $g^q_\sigma$ is the light-quark-$\sigma$ coupling constant, 
and $\psi_q$ is the light-quark wave function in the nucleon $N$ or 
hyperon $Y$ immersed in a nuclear medium.
By the above relation, we define explicitly the $\sigma$-$N$ and 
$\sigma$-$Y$ coupling constants:
\begin{equation}
g^{N,Y}_\sigma \equiv g_\sigma^{N,Y}(\sigma = 0) \equiv n_q g^q_\sigma S_{N,Y} (\sigma = 0). 
\label{sigmacc}
\end{equation}
Note that, the right hand side of Eq.~(\ref{Ssigma}) is the quark scalar charge, 
which is Lorentz scalar, and thus the left-hand-side of Eq.~(\ref{Ssigma}) 
is Lorentz scalar, and thus $m^*_N(\sigma)$ as well.
Furthermore, the values of $S_N(\sigma)$ and $S_Y(\sigma)$ are 
different, because the light-quark wave functions in the nucleon $N$ and hyperon $Y$ 
are different in vacuum as well as in medium, because the bag radii of the $N$ and $Y$ 
are different in each case.
Since the light quarks in the other hadrons feel the same scalar 
and vector mean fields as those in the nucleon, 
we can systematically study the hadron properties in medium 
without introducing any new coupling constants for   
the $\sigma$, $\omega$, and $\rho$ fields for different hadrons.

The parameters appearing at the nucleon, hyperon and meson Lagrangian level 
are $m_\omega = 783$ MeV, $m_\rho = 770$ MeV, $m_\sigma = 550$ MeV 
and $e^2/4\pi = 1/137.036$ ~\cite{Guichon:1995ue,Saito:1996sf}.
(See Ref.~\cite{Saito:1996sf} for a discussion on the parameter fixing in the QMC model, 
in treating finite nuclei.)

\section{Baryon properties in a nuclear medium}
\label{matter}

We consider the rest frame of infinitely large,  
symmetric nuclear matter, a spin and isospin saturated system 
with only strong interaction (Coulomb force is dropped as usual).
One first keeps only ${\cal L}^N_{QMC}$ in Eq.~(\ref{eq:LagYQMC}), 
or correspondingly drops all the quantities with the super- and sub-scripts $Y$, 
and sets the Coulomb field $A(\vec{r})=0$ in Eqs.~(\ref{eqdiracn})-(\ref{eqcoulomb}). 
Next one sets all the terms with any derivatives of the fields to be zero.  
Then, within the Hartree mean-field approximation, 
the nuclear (baryon) $\rho_B$ and scalar $\rho_s$ densities 
with the nucleon Fermi momentum $k_F$ are respectively given by,
\begin{eqnarray}
\hspace{-3ex} 
\rho_B &=& \dfrac{4}{(2\pi)^3}\int d^3{k}\ \theta (k_F - |\vec{k}|)
= \dfrac{2 k_F^3}{3\pi^2},
\label{rhoB}
\\
\hspace{-3ex}
\rho_s &=& \dfrac{4}{(2\pi)^3}\int d^3 {k} \ \theta (k_F - |\vec{k}|)
\dfrac{m_N^*(\sigma)}{\sqrt{m_N^{* 2}(\sigma)+\vec{k}^2}} .
\label{rhos}
\end{eqnarray}
Here, $m^*_N(\sigma)$ is the value (constant) of the effective nucleon mass at 
a given nuclear density.
In the standard QMC model~\cite{Guichon:1987jp}, the MIT bag model is used 
for describing nucleons and hyperons (hadrons). The use of this quark model is 
an essential ingredient for the QMC model, namely the use of the relativistic, 
confined quarks.

The Dirac equations for the quarks and antiquarks with the effective light-quark masses $m^*_q$ 
(to be defined below) 
in nuclear matter in a bag of a hadron $h$, with $q = u$ or $d$, and $Q = s,c$ or $b$,   
neglecting the Coulomb force are given 
by ~\cite{Tsushima:1997df,Tsushima:1998ru,Sibirtsev:1999jr,Tsushima:2002cc,Sibirtsev:1999js}, 
%
\begin{eqnarray}
& &\hspace{-3ex}
\left[ i \gamma \cdot \partial_x -
m^*_q
\mp \gamma^0
\left( V^q_\omega +
\dfrac{1}{2} V^q_\rho
\right) \right] 
\left( \begin{array}{c} \psi_u(x)  \\
\psi_{\overline{u}}(x) \\ \end{array} \right) = 0,
\label{Diracu}\\
& &\hspace{-3ex} 
\left[ i \gamma \cdot \partial_x -
m^*_q
\mp \gamma^0
\left( V^q_\omega -
\dfrac{1}{2} V^q_\rho
\right) \right]
\left( \begin{array}{c} \psi_d(x)  \\
\psi_{\overline{d}}(x) \\ \end{array} \right) = 0,
\label{Diracd}\\
& &\hspace{-3ex} 
\left[ i \gamma \cdot \partial_x - m_{Q} \right] \psi_{Q, \overline{Q}} (x) = 0, 
\label{DiracQ}
\end{eqnarray}
where, $m_q^* = m_q - V^q_\sigma$, and the (constant) mean fields for 
a bag in nuclear matter
are defined by $V^q_\sigma \equiv g^q_\sigma \sigma$, 
$V^q_\omega \equiv g^q_\omega \omega$ and
$V^q_\rho \equiv g^q_\rho b$,
with $g^q_\sigma$, $g^q_\omega$ and
$g^q_\rho$ being the corresponding quark-meson coupling constants. 
We assume SU(2) symmetry, $m_{u,\overline{u}}=m_{d,\overline{d}} \equiv m_q$, thus, 
$m^*_{u,\overline{u}}=m^*_{d,\overline{d}}=m^*_q 
\equiv m_q-V^q_{\sigma}$.
Since the $\rho$-meson mean field becomes zero, $V^q_{\rho}=0$~in Eqs.~(\ref{Diracu}) 
and~(\ref{Diracd}) in symmetric nuclear matter in the Hartree approximation,      
we will ignore it. (This is not true in a finite nucleus 
with equal and more than two protons  
even with equal numbers of protons and neutrons, 
since the Coulomb interactions among the protons induce an asymmetry  
between the proton and neutron density distributions 
to give $\rho_3(\vec{r}) = \rho_p(\vec{r}) - \rho_n(\vec{r}) \ne 0$.)

The same meson-mean fields $\sigma$ and $\omega$ for the quarks  
in Eqs.~(\ref{Diracu}) and~(\ref{Diracd}), satisfy self-consistently 
the following equations at the nucleon level, together with the effective 
nucleon mass $m_N^*(\sigma)$ of Eq.~(\ref{effnmass}) 
to be calculated by Eq.~(\ref{hmass}):
\begin{eqnarray}
{\omega}&=&\dfrac{g_\omega}{m_\omega^2} \rho_B,
\label{omgf}\\
{\sigma}&=&\dfrac{g^N_\sigma }{m_\sigma^2}C_N({\sigma}) \nn \\
& &\hspace{-3ex} \times \dfrac{4}{(2\pi)^3}\int d^3{k} \ \theta (k_F - |\vec{k}|)
\dfrac{m_N^*(\sigma)}{\sqrt{m_N^{* 2}(\sigma)+\vec{k}^2}}. 
\label{sigf} 
\end{eqnarray}
(See Eq.~(\ref{Ssigma}) for $C_N(\sigma)$.)
Because of the underlying quark structure of the nucleon to calculate
$m^*_N(\sigma)$ in nuclear medium, $C_N(\sigma)$ decreases as $\sigma$ increases,
whereas in the usual point-like nucleon-based models it is constant, $C_N(\sigma) = 1$. 
As will be discussed later it can be parametrized in the QMC model as 
$C_N(\sigma) = 1- a_N \times (g^N_\sigma \sigma)\,  (a_N > 0)$.
It is this variation of $C_N(\sigma)$ (or equivalently dependence of the scalar coupling on density,
or $\sigma$ as $g^N_\sigma (\sigma)$) that yields a novel saturation mechanism for nuclear matter 
in the QMC model, and contains the important dynamics originating from the quark structure
of nucleons and hadrons. It is also the variation of this $C_N(\sigma)$, 
that induces 3-body and higher order N-body forces~\cite{Guichon:2004xg}.
(This issue will be discussed separately in the next section.)
As a consequence of the {\em derived}, nonlinear
couplings of the meson fields in the Lagrangian density at the nucleon (hyperon) and meson level,
the standard QMC model yields the nuclear incompressibility of $K \simeq 280$~MeV. 
This is in contrast to a naive version of 
QHD~\cite{Walecka:1974qa,Serot:1984ey}
(the point-like nucleon model of nuclear matter),  
which results in the much larger value, $K \simeq 500$~MeV;
the empirically extracted value falls in the range $K = 200 - 300$ MeV.
(See Ref.~\cite{Dutra:2012mb} for an extensive analysis on this issue.)

\begin{figure}[htb]
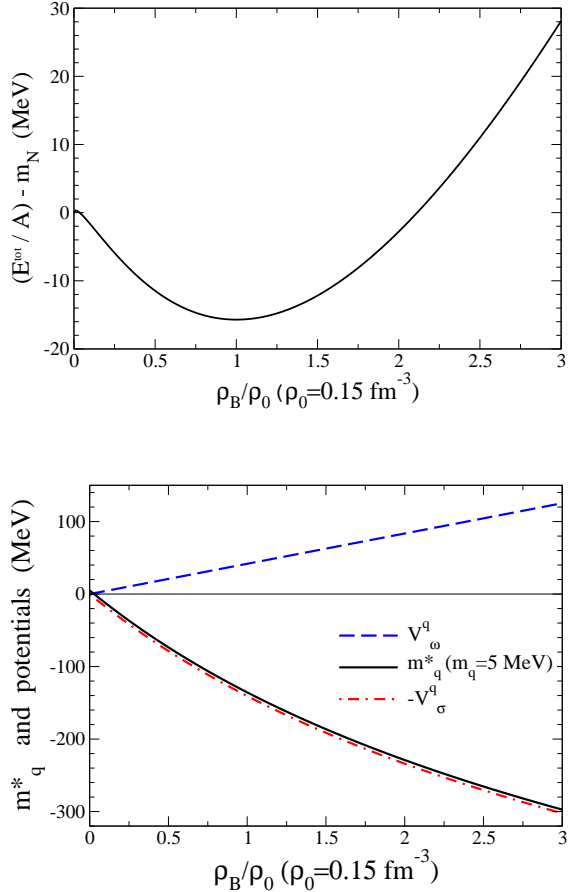

\vspace*{4ex}
\centering
\includegraphics[scale=0.3]{matterbcEnergyDen.eps}
\\
\vspace*{6ex}
\includegraphics[scale=0.3]{mqpot.eps}
\caption{\label{fig:Edenpot}
Negative of binding energy per nucleon for symmetric nuclear matter 
$E^{\rm tot}/A - m_N$ (upper panel), and 
the effective light-quark mass $m^*_q$, and vector ($V^q_\omega$) and 
scalar ($-V^q_\sigma$) potentials felt by the light quarks (lower panel).
}
\end{figure}

Once the self-consistency equation for the ${\sigma}$ field 
Eq.~(\ref{sigf}) is solved, one can evaluate the total energy of symmetric nuclear matter 
per nucleon:
\bg
& & E^{\rm tot}/A = \dfrac{4}{(2\pi)^3 \rho_B}\int d^3{k} 
\theta (k_F - |\vec{k}|) \sqrt{m_N^{* 2}(\sigma)+
\vec{k}^2} \nn \\
& &\hspace{8ex} +\dfrac{m_\sigma^2 {\sigma}^2}{2 \rho_B}+
\dfrac{g_\omega^2 \rho_B}{2m_\omega^2}.
\label{toten}
\en
We then determine the coupling constants, $g^N_\sigma$ 
and $g_{\omega}$ at the nucleon level (see also Eq.~(\ref{sigmacc})), by 
the fit to the binding energy of 15.7~MeV at the saturation density $\rho_0$ = 0.15 fm$^{-3}$
for symmetric nuclear matter, as well as  
$g_\rho$ to the symmetry energy of~35 MeV.
The determined quark-meson coupling constants, and the current quark mass values 
used are listed in Table~\ref{coupcc}.
The coupling constants at the nucleon level are  
$(g^N_\sigma)^2/4\pi = 3.12$, $g^2_\omega/4\pi = 5.31$ and $g^2_\rho/4\pi = 6.93$.
(See Eq.~(\ref{sigmacc}), and recall $g_\omega = 3 g^q_\omega$ and $g_\rho = g^q_\rho$.)
These values are determined with the standard QMC model inputs at the quark level 
which will be given later.

%
\begin{table}[htb]
\begin{center}
\caption{
Current quark mass values (inputs), quark-meson coupling constants 
and the bag pressure, $B_p$. Note that the $m_c$ value is 
updated from Refs.~\cite{Saito:2005rv,Krein:2017usp} 
based on Ref.~\cite{PDG}. 
}
\label{coupcc}
\vspace{1ex}
\begin{tabular}[t]{r|r||l|l}
\hline
\hline
$m_{u,d}$ &5    MeV &$g^q_\sigma$ &5.69\\
$m_s$     &250  MeV &$g^q_\omega$ &2.72\\
$m_c$     &1270 MeV &$g^q_\rho$   &9.33\\
$m_b$     &4200 MeV &$B_p^{1/4}$    &170 MeV\\
\hline
\hline
\end{tabular}
\end{center}
\end{table}

We show in Fig.~\ref{fig:Edenpot} negative of binding energy 
per nucleon for symmetric nuclear matter $E^{\rm tot}/A - m_N$ (upper panel), 
and effective light-quark mass $m^*_q$, 
vector ($V^q_\omega$) and scalar ($-V^q_\sigma$) potentials 
felt by the light quarks (lower panel).

Let us consider the situation that a hadron $h$ is immersed 
in nuclear matter. The normalized, static solution for the ground state quarks or antiquarks
with flavor $f$ in the hadron $h$ may be written,  
$\psi_f (x) = N_f \exp^{- i \epsilon_f t / R_h^*} \psi_f (\vec{r})$,
where $N_f$ and $\psi_f(\vec{r})$ are the normalization factor and
corresponding spin and spatial part of the wave function. 
The bag radius in medium for the hadron $h$, denoted by $R_h^*$, 
is determined through the
stability condition for the mass of the hadron against the
variation of the bag 
radius~\cite{Guichon:1987jp,Tsushima:1997rd}
(see Eq.~(\ref{stability})). 
The eigenenergies in units of $1/R_h^*$ are given by, 
\bg
\left( \begin{array}{c}
\epsilon_u \\
\epsilon_{\overline{u}}
\end{array} \right)
&=& \Omega_q^* \pm R_h^* \left(
V^q_\omega
+ \dfrac{1}{2} V^q_\rho \right),\,\,
\\
\left( \begin{array}{c} \epsilon_d \\
\epsilon_{\overline{d}}
\end{array} \right)
&=& \Omega_q^* \pm R_h^* \left(
V^q_\omega
- \dfrac{1}{2} V^q_\rho \right), \\
\epsilon_{Q}
&=& \epsilon_{\Qbar} =
\Omega_{Q}.
\label{energy}
\en

\noindent
The hadron mass in a nuclear medium, $m^*_h$ (free mass is denoted by $m_h$), is calculated 
for a given baryon density together with the mass stability condition,  
\begin{eqnarray}
& &m_h^* = \sum_{j=q,\overline{q},Q,\Qbar} 
\dfrac{ n_j\Omega_j^* - z_h}{R_h^*}
+ \frac{4}{3}\pi R_h^{* 3} B_p, 
\label{hmass} \\
& & \dfrac{d m_h^*}{d R^*_h} = 0,
\label{stability}
\end{eqnarray}
where $\Omega_q^*=\Omega_{\overline{q}}^*
=[x_q^2 + (R_h^* m_q^*)^2]^{1/2}\,(q=u,d)$, with
$m_q^* = m_q - g^q_\sigma \sigma=m_q-V^q_\sigma$,
$\Omega_Q^*=\Omega_{\Qbar}^*=[x_Q^2 + (R_h^* m_Q)^2]^{1/2}\,(Q=s,c,b)$,
and $x_{q,Q}$ are the lowest mode bag eigenvalues.
$B_p$ is the bag pressure (constant), $n_q (n_{\qbar})$ and $n_Q (n_{\Qbar})$ 
are the lowest mode valence quark (antiquark) 
numbers for the quark flavors $q$ and $Q$ 
in the hadron $h$, respectively, 
while $z_h$ parametrizes the sum of the
center-of-mass and gluon fluctuation effects,   
which are assumed to be density independent~\cite{Guichon:1995ue}. 
The bag pressure $B_p = {\rm (170\, MeV)}^4$ (density independent) is determined 
by the free nucleon mass 
$m_N = 939$ MeV with the bag radius in vacuum $R_N = 0.8$ fm and $m_q = 5$ MeV as inputs
(this yields $S_N(0) = 0.48265$ for Eq.~(\ref{sigmacc})), 
which are considered to be standard values in the QMC model~\cite{Saito:2005rv}.
(See also Table~\ref{coupcc}.)
Concerning the effective light-quark mass $m_q^*$ in nuclear 
medium, it reflects nothing but the strength 
of the attractive scalar potential 
as in Eqs.~(\ref{Diracu}) and~(\ref{Diracd}), 
and thus naive interpretation of the mass for a (physical) particle, 
which is positive, should not be applied. 
The model parameters are determined to reproduce the corresponding masses in free space.
The quark-meson coupling constants, $g^q_\sigma$, $g^q_\omega$
and $g^q_\rho$, have already been determined by the nuclear matter saturation properties.
Exactly the same coupling constants, $g^q_\sigma$, $g^q_\omega$, and
$g^q_\rho$ are used for the light quarks in all the hadrons  
as in the nucleon.

\begin{figure}[htb]
\centering
\includegraphics[scale=0.35,angle=-90]{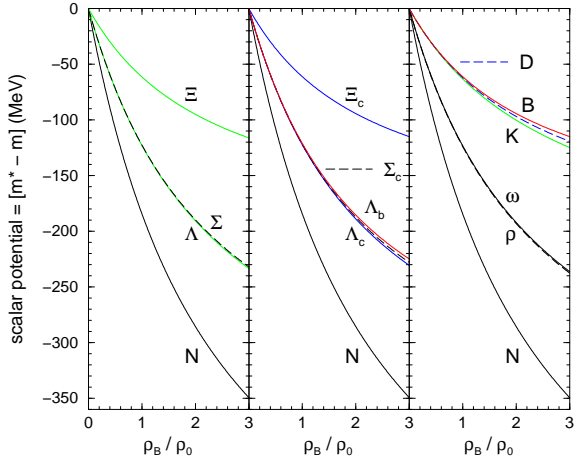}
\caption{\label{fig:scalar_pot} 
Baryon and meson scalar potentials, [$m^* - m$] (MeV)~\cite{Tsushima:2002cc}.
}
\end{figure}

We show in Fig.~\ref{fig:scalar_pot} the scalar potentials of 
baryons and mesons, $[m^* - m]$ (MeV), 
calculated in the QMC model~\cite{Tsushima:2002cc}.
(See Eq.~(\ref{hmass}) for $m^*$.)
One can notice that the scalar potentials of hadrons are  
well proportional to the light quark numbers of the corresponding hadrons.

In connection with the effective baryon masses,
it is found that the function $C_B({\sigma})$  
$(B = N,\Lambda,\Sigma,\Xi,\Lambda_c,\Sigma_c,\Xi_c,\Lambda_b,\Sigma_b,\Xi_b)$
(see Eq.~(\ref{Ssigma})), can be parameterized as a linear
form in the $\sigma$ field, $g^N_{\sigma}{\sigma}$, for a practical
use~\cite{Guichon:1995ue,Saito:1996sf,Tsushima:1997cu,Tsushima:2018goq}:
\bg
& & C_B ({\sigma}) = 1 - a_B
\times (g^N_{\sigma} {\sigma}),
\\
& & \hspace{8ex} (B = N,\Lambda,\Sigma,\Xi,\Lambda_c,\Sigma_c,\Xi_c,\Lambda_b,\Sigma_b,\Xi_b).
\nn
\label{cynsigma}
\en
The values obtained for $a_B$ are listed in Table~\ref{slope}.
This parameterization works well up to
about three times the normal nuclear matter density $3 \rho_0$.
Then, the effective mass of baryons $B$ in nuclear matter
is also well approximated up to $3 \rho_0$ by:
\bg
m^*_B &\simeq& m_B - \dfrac{n_q}{3} g^N_\sigma 
\left[1-\dfrac{a_B}{2}(g^N_\sigma {\sigma})\right]\sigma,
\\
& & = m_B - \dfrac{n_q}{3} \left[ g^N_\sigma \sigma - \dfrac{a_B}{2} (g^N_\sigma \sigma)^2 \right], 
\label{MBmass}
\\
& & \hspace{5ex}(B = N,\Lambda,\Sigma,\Xi,\Lambda_c,\Sigma_c,\Xi_c,\Lambda_b,\Sigma_b,\Xi_b),
\nn
\en
with $n_q$ being the valence light-quark number in the baryon $B$.
See Eqs.~(\ref{effnmass}) and~(\ref{effymass}) to compare with 
$g^{N,Y}(\sigma)$ and the above expression.
The obtained values of the ``slope parameter'' $a_B$ 
for various baryons are listed in Table~\ref{slope}.

%
\begin{table}[htb]
\begin{center}
\caption{Slope parameter values $a_B$ obtained for various 
baryons~\cite{Tsushima:2018goq}.
Note that the tiny differences in values of $a_B$ from those in  
Refs.~\cite{Saito:2005rv,Krein:2017usp}, are due to the differences 
in the number of data points for evaluating $a_B$, 
but such differences give negligible effects.   
}
\label{slope}
\begin{tabular}[t]{c|c||c|c||c|c}
\hline
$a_B$ &$\times 10^{-4}$ MeV$^{-1}$ &$a_B$ &$\times 10^{-4}$ MeV$^{-1}$ 
&$a_B$ &$\times 10^{-4}$ MeV$^{-1}$\\
\hline
$a_N$         &9.1  &--- &---                &--- &--- \\
$a_\Lambda$   &9.3  &$a_{\Lambda_c}$ &9.9    &$a_{\Lambda_b}$ &10.8 \\
$a_{\Sigma}$  &9.6  &$a_{\Sigma_c}$  &10.3   &$a_{\Sigma_b}$  &11.2 \\
$a_{\Xi}$     &9.5  &$a_{\Xi_c}$     &10.0   &$a_{\Xi_b}$     &10.8 \\
\hline
\end{tabular}
\end{center}
\end{table}
%

%
\section{The QMC model and conventional nuclear models
\label{sec:conventional}}
%

In this section we discuss the relationship between the QMC model and 
a conventional Skyrme effective nuclear force  
according to Ref.~\cite{Guichon:2004xg}. (For a review including further developments, 
see Refs.~\cite{Guichon:2006er,Stone:2016qmi,Guichon:2018uew,Stone:2019syx}.)
The QMC model description was reformulated to describe 
a nucleus as a many-body problem in a nonrelativistic framework. 
This allows us to take the limit corresponding  
to a zero-range force which can be compared with the Skyrme 
effective forces in conventional nuclear physics~\cite{Guichon:2004xg}. 

The classical energy of a nucleon with position ($\vec{r}$) 
and momentum ($\vec{p}$) is given by~\cite{Guichon:2004xg}, 
\begin{equation}
E_{N}(\vec{r}) =  \frac{\vec{p}^{\,2}}{2m_N^\ast(\vec{r})}+m_N^\ast(\vec{r})+
g_{\omega }\omega (\vec{r})+V_{s.o.},  
\label{Eq-QMC1}
\end{equation}
where $V_{s.o.}$ is the spin-orbit interaction.

To get the dynamical mass $m_N^\ast(\vec{r})$
one has to solve a quark model of the nucleon (in the present case the MIT bag model) 
in the field $\sigma (\vec{r})$.
For the present purpose, it is sufficient to use the approximated relation 
Eq.~(\ref{MBmass}) with $n_q = 3$ and $d = a_N$ and $g_\sigma \equiv g^N_\sigma$ hereafter,
\begin{equation}
m_N^\ast(\vec{r}) = m_N-g_{\sigma }\sigma (\vec{r})+
\frac{d}{2}\left( g_{\sigma }\sigma (\vec{r})\right)^{2},  
\label{Eq-QMC4}
\end{equation}
where $d$ of the MIT bag model gives $d=0.22 R_{N}$ (in MeV$^{-1}$) with 
the nucleon bag radius $R_N$ (fm) corresponding 
to Table~\ref{slope} with $R_N = 0.8$ fm. 
The last term, which represents the response of the
nucleon to the applied scalar field -- the scalar polarizability --    
is an essential element of the QMC model. From
the numerical studies we know that the approximation 
Eq.~(\ref{Eq-QMC4}) 
is quite accurate at moderate nuclear densities.  

The energy~(\ref{Eq-QMC1}) is for one particular nucleon
moving classically in the nuclear meson fields. 
The total energy of the system is then given by the sum
of the energy of each nucleon and the energy carried by
the fields~\cite{Saito:2005rv}:
\begin{eqnarray}
E_{tot}&=&\sum _{i}E_{N}(\vec{r}_{i})+E_{meson},  \\
E_{meson} &=& \frac{1}{2}\int d^3 r\, [
\left( {\vec \nabla}\sigma \right)^{2} + 
m_{\sigma }^{2}\sigma^{2} \nn \\ 
& &\hspace{12ex} -\left( {\vec \nabla}\omega 
\right)^{2}-m_{\omega }^{2}\omega^{2} ]. 
\label{mesonenergy} 
\end{eqnarray}

The expression of $E_{N}(\vec{r})$ was approximated 
by neglecting the velocity dependent terms 
$( {\vec \nabla}\sigma )^{2}$, 
\begin{eqnarray}
E_{tot} &=& E_{meson}+\sum _{i}\left( m_N +
\frac{\vec{p}^{\,2}_{i}}{2m_N}+V_{so}(i)\right) \nn \\
& & - \int d^3 r\, \rho ^{cl}_{s}\, \left( 
g_{\sigma }\sigma -\frac{d}{2}(g_{\sigma }\sigma )^{2}\right) \nn \\ 
& & + \int d^3 r\, \rho ^{cl}\, g_{\omega }\omega , 
\label{Eq-QMC10} 
\end{eqnarray}
where we define the classical densities as $\rho^{cl}(\vec{r})=\sum _{i}\delta 
(\vec{r}-\vec{r}_{i})$ and 
$\rho^{cl}_{s}(\vec{r})=\sum _{i}(1-{\vec{p}^{\,2}_{i}}/2m_N^{2}) 
\delta (\vec{r}-\vec{r}_{i})$. 
This will be the starting point for the many body formulation of
the QMC model. 

To eliminate the meson fields from the energy, we use the equations, 
$\delta E_{tot}/\delta \sigma (\vec{r})=\delta E_{tot}/\delta \omega (\vec{r})=0$, 
and leave a system whose dynamics depends only on the nucleon coordinates. 
Roughly speaking, since the meson fields should follow
the matter density, the typical scale for the ${\vec \nabla}$ 
operator acting on $\sigma$ or $\omega$ is the thickness 
of the nuclear surface, that is about $1$ fm. 
Therefore, it seems reasonable that we can consider the second derivative 
terms acting on the meson fields as perturbations. 
Then, starting from the lowest order approximation, we solve the equations for the 
meson fields iteratively, and neglect a small difference between 
$\rho^{cl}_{s}$ and $\rho^{cl}$ except in the leading term. 
When inserted into Eq.~(\ref{Eq-QMC10}), the series for the meson fields generates 
$N$-body forces in the Hamiltonian. 
To complete the effective Hamiltonian, we now include the effect of
the isovector $\rho$ meson as well.  

The quantum effective Hamiltonian finally takes the form 
\begin{eqnarray}
 H_{QMC} &=& \sum _{i}\frac{{\overleftarrow  \nabla}_{i}\cdot 
 {\overrightarrow \nabla}_{i}}{2m_N}+
 \frac{G_{\sigma }}{2m_N^{2}}\sum _{i\neq j}\overleftarrow{\nabla_{i}}
 \delta (\vec{r}_{ij})\cdot \overrightarrow{\nabla}_{i} 
\nn\\ 
& & +\frac{1}{2}\sum _{i\neq j}\left[ {\vec \nabla}^{2}_{i}\delta (\vec{r}_{ij})\right] 
\left( \frac{G_{\omega }}{m_{\omega }^{2}}-\frac{G_{\sigma }}{m_{\sigma }^{2}}
+\frac{G_{\rho }}{m_{\rho }^{2}}\frac{\vec{\tau }_{i}.\vec{\tau }_{j}}{4} \right) 
\nn\\
& &+ \frac{1}{2}\sum _{i\neq j}\delta (\vec{r}_{ij})\left[ G_{\omega }-G_{\sigma }
+G_{\rho }\frac{\vec{\tau }_{i}.\vec{\tau }_{j}}{4}\right] 
\nn\\
& & +\frac{dG_{\sigma }^{2}}{2}\sum _{i\neq j\neq k}\delta^{2}(ijk)
-\frac{d^{2}G_{\sigma }^{3}}{2}\sum _{i\neq j\neq k\neq l}\delta ^{3}(ijkl) 
\nn\\
& & + \frac{i}{4m_N^{2}}\sum _{i\neq j}A_{ij}\overleftarrow{\nabla}_{i}
\delta (\vec{r}_{ij})\times \overrightarrow{\nabla}_{i}\cdot \vec{\sigma }_{i} ,
\label{Eq-QMC28} 
\end{eqnarray}
where $G_i=g_i^{2}/m_i^{2}$ ($i= \sigma, \omega, \rho$) and $A_{ij}=G_{\sigma }+(2\mu 
_{s}-1)G_{\omega }+
(2\mu _{v}-1)G_{\rho }\vec{\tau }_{i}\cdot \vec{\tau }_{j}/4$, 
with $\mu_s$ and $\mu_v$ being respectively, the nucleon isoscalar and isovector magnetic moments.
Here $\vec{r}_{ij}=\vec{r}_{i}-\vec{r}_{j}$ and ${\vec \nabla}_{i}$ 
is the gradient with respect to $\vec{r}_{i}$ acting on the delta
function. In Eq.~(\ref{Eq-QMC28}) we have used the notation
$\delta^{2}(ijk)$ for $\delta (\vec{r}_{ij})\delta (\vec{r}_{jk})$ 
and analogously for $\delta ^{3}(ijkl)$. Furthermore, we have dropped the contact interactions
involving more than 4-bodies because their matrix elements vanish
for antisymmetrized states. 

To fix the free parameters, $G_i$, 
the volume and symmetry coefficients of 
the binding energy per nucleon of infinite
nuclear matter, $E_{B}/A=a_{1}+a_{4}(N-Z)^{2}/A^{2}$, are calculated and fitted so as to produce 
the 
experimental values. Using the bag model with the radius 
$R_N=0.8$ fm and the physical masses for the mesons and 
$m_{\sigma}= 600$ MeV, one gets, in fm$^{2}$,  
$G_{\sigma }=11.97$, $G_{\omega }=8.1$ and $G_{\rho }=6.46$.

It is now possible to compare the present Hamiltonian with the Skyrme effective interaction. 
Since, in our formulation, the medium effects are summarized in the 3- and 4-body forces, 
we consider Skyrme forces of the same type, that is, without density dependent interactions. They 
are defined by a potential energy of the form 
\begin{eqnarray}
V &=& t_{3}\sum_{i<j<k}\delta (\vec{r}_{ij})\delta (\vec{r}_{jk})
\nn\\
& & + \sum_{i<j} \left[ t_{0}(1+x_{0}P_{\sigma})\delta (\vec{r}_{ij}) 
\right. \nn\\
& & +\frac{1}{4}t_{2}\, \overleftarrow{\nabla }_{ij}\cdot \delta 
(\vec{r}_{ij})\overrightarrow{\nabla }_{ij} 
\nn\\
& & - \frac{1}{8}t_{1}\left( \delta (\vec{r}_{ij}){\overrightarrow  \nabla}_{ij}^{2}+
\overleftarrow{\nabla}_{ij}^{2}\delta (\vec{r}_{ij}) \right)
\nn\\
& & \left. +\frac{i}{4}W_{0}(\vec{\sigma }_{i}+\vec{\sigma }_{j})\cdot 
\overleftarrow{\nabla}_{ij}
\times \delta (\vec{r}_{ij})\overrightarrow{\nabla}_{ij}^2  \right],
\label{Eq-QMC53} 
\end{eqnarray}
with $\nabla_{ij}=\nabla_{i}-\nabla_{j}$. There is no 4-body
force in Eq.~(\ref{Eq-QMC53}).  Comparison of Eq.~(\ref{Eq-QMC53})
with the QMC Hamiltonian, Eq.~(\ref{Eq-QMC28}), allows one to identify
\begin{equation}
t_{0}=-G_{\sigma }+G_{\omega }-\frac{G_{\rho }}{4}, \ \ \ 
t_{3}=3dG_{\sigma }^{2}, \ \ \ x_{0}=-\frac{G_{\rho }}{2t_{0}}  .
\label{Eq-QMC100}
\end{equation} 

Furthermore, we restrict our considerations to doubly
closed shell nuclei, and assume that one can
neglect the difference between the radial wave functions of 
the single-particle states with $j=l+1/2$ and $j=l-1/2$.  
Then, by comparing the Hartree-Fock Hamiltonian obtained
from $H_{QMC}$ and that of Ref.~\cite{VAU} corresponding to the Skyrme force, 
we obtain the relations 
\begin{eqnarray}
3t_{1}+5t_{2}&=&\frac{8G_{\sigma }}{m_N^{2}}+4\left( \frac{G_{\omega }}{m_{\omega }^{2}}
-\frac{G_{\sigma }}{m_{\sigma }^{2}}\right)
  +3\frac{G_{\rho }}{m_{\rho }^{2}} ,\label{EqQMC110}\\
5t_{2}-9t_{1}&=&\frac{2G_{\sigma }}{m_N^{2}}+28\left( \frac{G_{\omega }}{m_{\omega }^{2}}
-\frac{G_{\sigma }}{m_{\sigma
 }^{2}}\right) -3\frac{G_{\rho }}{m_{\rho }^{2}} , \label{5t2}\\
W_{0} &=& \frac{1}{12m_N^{2}} ( 5G_{\sigma }+5(2\mu _{s}-1)G_{\omega } 
\nn\\
& & \hspace{12ex} + \frac{3}{4}(2\mu _{v}-1)G_{\rho } ) . \label{w0}
\end{eqnarray}

We compare in Table~\ref{tab:Tab-QMB4} the results with the parameters
of the force SkIII~\cite{FRI}, which is considered a good representative 
of density independent effective interactions. 
We show the combinations $3t_{1}+5t_{2}$,
which controls the effective mass, and $5t_{2}-9t_{1}$, which controls
the shape of the nuclear surface~\cite{VAU}.  
From the Table~\ref{tab:Tab-QMB4}, one sees that the
level of agreement with SkIII is very impressive. 
An important point is that the spin-orbit strength 
$W_{0}$ comes out with approximately the correct value. 
The middle column (N=3) shows the results when
we switch off the 4-body force. 
The main change is expected to decrease of the predicted 3-body force. 
However, this is not the case. If we look at the incompressibility of nuclear 
matter, $K$, this decreases by as much as $37$ MeV when we restore this 4-body force. 

Now one can recognize a remarkable agreement between the
phenomenologically successful Skyrme force (SkIII) 
and the effective interaction
corresponding to the QMC model --- a result which  
suggests that the response of nucleon internal structure to the nuclear
medium ($scalar polarisability$) indeed plays a vital role in nuclear structure.

\begin{table}[htb]
\begin{center}
\caption{QMC predictions (with $m_\sigma =600$ MeV)~\cite{Guichon:2004xg} 
compared with the Skyrme force~\cite{FRI}.}
\label{tab:Tab-QMB4}
\begin{tabular}{c|c|c|c}
\hline
& QMC & QMC(N=3) & SkIII \\
\hline
$t_0$ (MeV\,fm$^3$) & -1082 & -1047 & -1129 \\
$x_0$ & 0.59 & 0.61 & 0.45 \\
$t_3$ (MeV\,fm$^6$) & 14926 & 12513 & 14000 \\
$3t_1+5t_2$ (MeV\,fm$^5$) & 475 & 451 & 710 \\
$5t_2-9t_1$ (MeV\,fm$^5$) & -4330 & -4036 & -4030 \\
$W_0$ (MeV\,fm$^5$) & 97 & 91 & 120 \\
$K$ (MeV) & 327 & 364 & 355 \\
\hline
\end{tabular}
\end{center}
\end{table}
%

\section{Summary}
\label{summary}

We have given a short review on the basics of the quark-meson coupling (QMC) 
model, a quark-based model of finite nuclei and hadron properties in a nuclear medium. 
The highlight was on the relationship between the QMC model and 
a conventional Skyrme effective nuclear force, by reformulating 
the QMC model in nonrelativistic form and taking the zero-range interaction limit.
It was shown that the derived, effective QMC interaction has a remarkable 
agreement with a successful Skyrme force.
Furthermore, it was shown that the QMC-generated effective interaction 
automatically contains the 3-body and higher order N-body forces.
Since the QMC model is based on the quark degrees of freedom,  
the model enables us to study the properties of finite nuclei and  
in-medium hadron properties in a very systematic manner.

\begin{acknowledgments}
The author would like to thank P.~A.~M.~Guichon, K.~Saito, and A.~W.~Thomas 
for exciting collaborations, and the Asia Pacific Center for Theoretical Physics (APCTP),   
Kyungpook National University, and Yongseok Oh for warm hospitality 
and supports during his visit.
This work was supported by the Conselho Nacional de Desenvolvimento 
Cient\'{i}fico e Tecnol\'{o}gico (CNPq) 
Process, No.~313063/2018-4, and No.~426150/2018-0, 
and Funda{\c c}\~{a}o de Amparo \`{a} Pesquisa do Estado 
de S\~{a}o Paulo (FAPESP) Process, No.~2019/00763-0, 
and was also part of the projects, Instituto Nacional de Ci\^{e}ncia e 
Tecnologia -- Nuclear Physics and Applications (INCT-FNA), Brazil, 
Process. No.~464898/2014-5, and FAPESP Tem\'{a}tico, Brazil, Process, 
No.~2017/05660-0.
\end{acknowledgments}



\end{document}